\documentclass[aps,prl,twocolumn,superscriptaddress,nofootinbib]{revtex4-2}

\usepackage{amsmath,amssymb,amsfonts}
\usepackage{xcolor, graphicx}
\usepackage{bm}

\begin{document}
\begin{flushright}
\small CERN-TH-2026-178
\end{flushright}

\title{Imaging Non-Hydrodynamic Modes with Jet Wakes}

\author{Aleksi Kurkela}
\affiliation{Department of Mathematics and Physics, University of Stavanger, 4036 Stavanger, Norway}

\author{Ian Moult}
\affiliation{Department of Physics, Yale University, New Haven, Connecticut, 06511, USA}

\author{Alexander Soloviev}
\affiliation{Faculty of Mathematics and Physics, University of Ljubljana, 1000 Ljubljana, Slovenia}

\author{Urs Achim Wiedemann}
\affiliation{Theoretical Physics Department, CERN, CH-1211 Geneva 23, Switzerland}

\date{\today}

\begin{abstract}
While studies of ultra-relativistic heavy-ion collisions have established that the quark--gluon plasma exhibits hydrodynamic behavior, direct signatures of non-hydrodynamic modes have remained elusive, and no observable is known to be exclusively sensitive to them. Here, we show that the angular structure of the jet wake provides such a probe. In the long-wavelength limit, hydrodynamics contributes only to the lowest angular moments of the detector image of the jet wake, while higher moments directly encode microscopic non-equilibrium dynamics. The jet wake thus serves as a spectroscopic probe of the medium's non-hydrodynamic sector. We develop a general kinetic-theory framework relating the angular moments of the late-time energy flux generated by a jet to the relaxation spectrum of the collision operator. In all models considered, non-hydrodynamic modes leave distinct imprints on the higher angular moments. Our results motivate precision measurements of the higher angular moments of the negative jet wake.
\end{abstract}

\maketitle

{\bf Introduction.}
The plasmas of all relativistic quantum field theories support both hydrodynamic and non-hydrodynamic excitations. Hydrodynamic modes --- including sound and the diffusive modes associated with shear, heat, and charge --- govern the long-wavelength, late-time dynamics. They are universal consequences of conservation laws and are fully governed by relativistic hydrodynamics~\cite{Baier:2007ix}. Plasmas also support non-hydrodynamic excitations~\cite{Kovtun:2005ev,Romatschke:2015gic}, whose microscopic nature depends on the underlying quantum field theory~\cite{Hong:2010at,Kurkela:2017xis,Moore:2018mma,Grozdanov:2019uhi,Ochsenfeld:2023wxz,Grozdanov:2024fxr}, ranging from quasi-normal modes in strongly coupled theories with gravity duals to quasi-particle excitations in weakly coupled theories. Experimental access to these modes may therefore provide direct information about both the microscopic structure of the plasma~\cite{Kurkela:2019kip,Ambrus:2021fej,Ambrus:2022koq} and the early-time dynamics of its approach to thermal equilibrium~\cite{Heller:2015dha,Heller:2016rtz}.

Ultra-relativistic heavy-ion collisions at RHIC and the LHC probe the properties of the quark--gluon plasma (QGP), the plasma of QCD, whose
hydrodynamic sector has been tightly constrained in a close interplay between experiment and theory. In particular, a broad body of long-wavelength (low-transverse-momentum) data is accurately described by relativistic fluid dynamics \cite{Bernhard:2019bmu, Nijs:2020ors}, yielding constraints on the QGP equation of state and transport coefficients, including the shear $\eta$ and bulk $\zeta$ viscosities governing the dissipative relaxation of conserved quantities. By contrast, comparable experimental constraints on the non-hydrodynamic sector remain absent.

Accessing non-hydrodynamic excitations requires driving the system away from equilibrium. Heavy-ion collisions naturally realize such conditions during the earliest stages, before hydrodynamization, and in small collision systems, such as light-ion or peripheral heavy-ion collisions, where the medium may not fully equilibrate. In practice, however, the subsequent hydrodynamic evolution of large systems erases sensitivity to early-time dynamics, while analyses of small systems remain limited by uncertainties in the initial state and hadronization. As a result, neither approach has so far yielded conclusive evidence for non-hydrodynamic excitations in the QGP.

Here, we argue that the medium response to jets~\cite{Casalderrey-Solana:2004fdk,Neufeld:2008fi} provides a uniquely controlled out-of-equilibrium probe that circumvents these limitations. Specifically, jets can excite modes of higher spherical harmonics, whereas hydrodynamic excitations contribute only to $\ell\leq 2$, with the $\ell=2$ mode related to the shear viscosity. Modes with $\ell>2$ therefore lie outside the hydrodynamic sector, making jet medium response a qualitatively novel spectroscopic probe of non-hydrodynamic excitations. We illustrate the relevance of this general idea by focusing on the negative jet wake --- the depletion of particle and energy flow opposite to the jet direction~\cite{Chen:2021gkj} associated with the redistribution of energy and momentum deposited by the jet --- which has recently been observed by CMS in $Z$-tagged jet events as well as in dijet events with a rapidity gap~\cite{CMS:2025dua}. In this Letter, we show concretely how the low-$\ell$ angular structure of the detector image generated by the jet wake encodes the non-hydrodynamic modes governing the relaxation of the medium. 

{\bf Sourced Boltzmann equation.} To demonstrate the sensitivity of the jet wake to non-hydrodynamic excitations, we employ a description of the collective QGP dynamics that extends beyond hydrodynamics and couples to jets. Specifically, we consider a generic kinetic theory for the phase distribution function $f$, 
\begin{equation}
p^\mu \partial_\mu \, f(x,\mathbf p)
= - p^0\,\mathcal C[f](x,\mathbf p)
+ p^0\,S(x,\mathbf p) \, ,
\label{boltzmann}
\end{equation}
where the source $S(x,\mathbf p)$ accounts for jet-medium interactions.  The collision operator
$\mathcal C$ encodes the microscopic interactions governing the relaxation of the medium. This framework encompasses relativistic hydrodynamics while also propagating non-hydrodynamic degrees of freedom. For example, in weakly coupled quantum field theories, the late-time, long-wavelength behavior of correlation functions is governed by the transport of on-shell quasiparticles~\cite{Jeon:1995zm,Arnold:2002zm} and can be described by a Boltzmann equation for QCD and other quantum field theories. 

Experimentally, the medium response to jets does not provide access to the full phase space distribution $f$, but only to its detector image,  which is asymptotic particle flow recorded at late times
\begin{equation}
 f_{\rm image}(\mathbf p)  =  \lim_{t\to\infty}
 \int d^3\mathbf x \, f(t,\mathbf x,\mathbf p).
 \label{image}
\end{equation}
We focus on calorimetric measurements of the jet-induced energy flux along a direction $\mathbf{\hat n}$
\begin{equation}
T^{0i}\hat n_i = \int_0^\infty \frac{dp \, p^3}{(2\pi)^3} \, f_{\rm image}(p\,\mathbf{\hat n}) .
\label{eflow}
\end{equation}
In ultra-relativistic heavy-ion collisions, the evolution of $f$ from an initially dense medium to a dilute one involves hadronization and non-perturbative processes that are difficult to model. However, energy is conserved throughout the evolution, making calorimetric measurements particularly robust against assumptions about late-time dynamics. The angular distribution \eqref{eflow} therefore shares the robustness that motivates recent studies of 
energy-energy correlators \cite{Moult:2025nhu}. 

We discuss jet-induced medium response in an idealized geometry. Until time $t_*$, the unperturbed medium is described by the local equilibrium distribution 
\begin{equation}
f_{\rm eq}(\mathbf p;T,\mathbf u) = f_0\!\left( \frac{p^0-\mathbf p\cdot\mathbf u}{T}
\right)\, ,
\label{equilibrium}
\end{equation}
parametrized by the local temperature $T$ and fluid velocity \(\mathbf u\). Freeze-out, i.e. the conversion of the QGP into hadrons, is assumed to occur instantaneously at time $t_*$, after which interactions are negligible for the energy flow in \eqref{eflow}.

The sourced Boltzmann equation \eqref{boltzmann} can be linearized about this equilibrium 
in terms of perturbations $|\delta f(\omega,\mathbf k)\rangle$ 
\begin{equation}
\left[ -i(\omega-\mathbf k\cdot\mathbf v_p)+\hat{\mathcal C} \right] |\delta f(\omega,\mathbf k)\rangle
= |S(\omega,\mathbf k)\rangle\, .
\end{equation}
Here \(K^\mu=(\omega,\mathbf k)\) denotes the Fourier-conjugate of $(t,{\bf x})$, \(\mathbf v_p=\mathbf p/p^0\) is the particle velocity, and the collision kernel \(\hat{\mathcal C}\) is the linearized collision operator.
Since every local equilibrium distribution is stationary, perturbations $  |\delta f(\omega,\mathbf k)\rangle = |\phi_a\rangle$ tangent to the equilibrium manifold, 
\begin{equation}
\langle \mathbf p|T\rangle = \frac{\partial f_{\rm eq}}{\partial T},
\qquad \langle \mathbf p|u_i\rangle = \frac{\partial f_{\rm eq}}{\partial u^i}\, ,
\end{equation}
are conserved. They span the null space of the collision kernel,
\begin{equation}
\hat{\mathcal C}|\phi_a\rangle =0,
\qquad
a\in {\rm cons}\, .
\end{equation}
Such conserved perturbations of \eqref{equilibrium} are the hydrodynamic modes.

In general, the space of perturbations $  |\delta f(\omega,\mathbf k)\rangle$ can be spanned by an orthonormal basis $\lbrace |\phi_a\rangle \rbrace$ of eigenmodes of the collision kernel,
\begin{equation}
\hat{\mathcal C}
= \sum_a \frac{1}{\tau_a} |\phi_a\rangle\langle\phi_a|,
\qquad \langle\phi_a|\phi_b\rangle=\delta_{ab}\, ,
\end{equation}
which is equipped with the kinetic theory inner product 
\begin{equation}
\langle \phi_a|\phi_b\rangle
= \int_{\mathbf p} w^{-1}(p)\,
\phi_a(\mathbf p)\phi_b(\mathbf p),
\qquad
w(p)\equiv -\frac{\partial f_0}{\partial p^0},
\end{equation}
with respect to which the collision kernel is self-adjoint. The eigenvalues \(1/\tau_a\) determine the relaxation rates of the corresponding excitations. Characteristic non-hydrodynamic modes,  such as the quasi-particles in weakly coupled field theories or the quasi-normal modes in theories with gravity duals, are characterized by relaxation times that can be extracted from the analyticity properties of retarded two-point energy correlation functions.

{\bf The detector state.}
The detector image can be expressed directly in terms of the eigenmodes of the collision kernel.
Because it is obtained from the spatially integrated distribution \eqref{image}, it is sensitive only to the zero-momentum sector, \(\mathbf k\to0\), yielding
\begin{equation}
|\delta f(\omega)\rangle = \left[ -i\omega+\hat{\mathcal C} \right]^{-1}
|S(\omega)\rangle\, .
\end{equation}
We characterize the resulting angular distribution of the late-time energy flux by its spherical-harmonic moments,
\begin{equation}
\delta \mathcal E_{\ell m} = \int d\Omega_{\hat{\mathbf n}}\, Y_{\ell m}^*(\hat{\mathbf n})\,
\frac{d\delta T^{0r}}{d\Omega}(\hat{\mathbf n})
\equiv
\langle I_{\ell m}|\delta f\rangle .
\end{equation}
The corresponding detector state is
\begin{equation}
\langle I_{\ell m}|\mathbf p\rangle
= w(p)\,p\, Y_{\ell m}^*(\hat{\mathbf p}) .
\end{equation}
Because the equilibrium background is rotationally invariant, the
collision operator commutes with rotations, and its eigenmodes can be
chosen to have definite angular momentum,
\begin{equation}
|\phi_a\rangle \equiv
|\phi_{n\ell m}\rangle,
\qquad
\langle \mathbf p|\phi_{n\ell m}\rangle
= \phi_{n\ell}(p)\, Y_{\ell m}(\hat{\mathbf p}).
\end{equation}
Here \(\ell,m\) characterize the angular structure of the excitation, while \(n\) labels independent radial modes within each angular channel. This decomposition cleanly separates different classes of excitations:  conserved modes correspond to \(n=0,1\), while non-conserved modes have \(n>1\). As we discuss below, the subspace $\ell \geq 3$ contains only non-hydrodynamic excitations. This motivates the study of detector moments, that are sensitive only to excitations in the corresponding angular channel,
\begin{equation}
\delta \mathcal E_{\ell m} = \sum_n \langle I_{\ell m}|\phi_{n\ell m}\rangle
\, \langle \phi_{n\ell m}|\delta f\rangle .
\end{equation}

Combining the detector decomposition with the spectral representation of
the collision kernel, the detector moments become
\begin{equation}
\delta\mathcal E_{\ell m}
=
\sum_n
\frac{
\langle I_{\ell m}|\phi_{n\ell m}\rangle
\langle\phi_{n\ell m}|S(\omega)\rangle
}{
-i\omega+\tau_{n\ell}^{-1}
},
\end{equation}
with relaxation times \(\tau_{n\ell}\).

{\bf Jet wake and its image.}
We specialize to a jet-induced source, modeled as an eikonal color charge propagating through the medium. The interaction with the plasma generates a nonlocal perturbation through two-gluon exchange. This source was previously discussed in coordinate space~\cite{Neufeld:2008fi} after integrating over the momentum dependence of the medium constituents. Here (see Appendix) we retain the momentum dependence and consider only the spatially integrated (\(\mathbf k\to0\)) component, which contributes to the detector image,
\begin{equation}
S(\mathbf p,\omega)
\equiv
\lim_{\mathbf k\to0}
\langle \mathbf p|S(\omega,\mathbf k)\rangle .
\end{equation}

We focus on jets evolving such
that the response is dominated by the low-frequency regime \((\omega\to0)\), and we denote the
corresponding momentum-space source by \(|S(\omega) \rangle = 2\pi \delta(\omega)|S\rangle\). Because the jet defines the \(z\)-axis, the source is azimuthally symmetric, and only the \(m=0\) angular components are excited. We
therefore suppress the \(m\)-index in the following.

Assuming the source remains approximately constant until freeze-out at
time \(t_*\), the detector moments are
\begin{equation}
\delta\mathcal E_{\ell} = \sum_n
\tau_{n\ell} \left( 1-e^{-t_*/\tau_{n\ell}} \right)
\langle I_{\ell}|\phi_{n\ell}\rangle
\langle \phi_{n\ell}|S\rangle .
\label{eq:image}
\end{equation}
Here, modes with $\tau_{n\ell} \gg t_*$ behave
effectively as conserved excitations, since
$\tau_{n\ell} \left( 1-e^{-t_*/\tau_{n\ell}} \right)
\to t_*  $.
In contrast, modes with $\tau_{n\ell} \lesssim t_*$ retain memory only over
times of order \(\tau_{n\ell}\) prior to freeze-out and therefore probe
the non-hydrodynamic relaxation spectrum of the medium.

Eq.~\eqref{eq:image} shows that different angular moments probe
different sectors of the kinetic response. The lowest angular channels have distinct physical interpretations: 

\emph{Conserved sector, $\ell=0,1$}:
The $\ell=0,1$ channels contain both conserved and non-conserved modes.
However, the monopole and dipole detector moments coincide with the
conserved energy and momentum modes
\begin{equation}
\langle I_{00}|=\langle T|, \qquad \langle I_{1m}| = \langle u_m|.
\end{equation}
and are therefore orthogonal to all non-conserved excitations,
\begin{equation}
\langle I_{00}|\phi_{n00}\rangle =0, \qquad \langle I_{1m}|\phi_{n1m}\rangle =0, \qquad n>0.
\end{equation}
Consequently, the monopole contribution is completely determined by the
energy deposited by the jet, while the dipole contribution is completely
determined by the momentum deposited by the jet,
\begin{equation} \delta \mathcal E_0 = t_* \langle T|S\rangle, \qquad \delta \mathcal E_1 = t_* \langle u_0|S\rangle .
\end{equation}
They therefore probe the energy and momentum deposited by the
jet and can be correlated with independent measurements of jet
energy loss. The $\ell=1$ mode describes the depletion of energy in the direction opposite to the jet, giving rise to the negative jet wake.

\emph{Shear sector, $\ell=2$}:
No conserved mode exists in the quadrupole sector $\ell=2$ where the detector moments receive contributions
from the full spectrum of non-conserved excitations,
\begin{equation}
\delta \mathcal E_2 = \sum_n \tau_{n2}\, \langle I_2|\phi_{n2}\rangle \langle \phi_{n2}|S\rangle .
\end{equation}
This sector probes the relaxation of elliptically deformed momentum
distributions and is therefore closely related to the shear viscosity,
which is governed by the same \(\ell=2\) relaxation spectrum,
\begin{equation}
\eta
=
\sum_n
\tau_{n2}\,
\langle S_{\rm shear}|\phi_{n2} \rangle
\langle \phi_{n2}|S_{\rm shear}\rangle .
\end{equation}
Here $
\langle \mathbf p|S_{\rm shear}\rangle
=
\sqrt{\frac{4\pi}{15}}\,
\frac{p^2}{p^0}\,
Y_{20}(\hat{\mathbf p})\,
w(p),
$ denotes the source generated by a velocity-gradient
perturbation obtained for the Kubo relations~\cite{Arnold:2003zc, Moore:2018mma}. In a purely hydrodynamic description, the quadrupole
response is completely determined by the viscous stress tensor, so that
the jet quadrupole image becomes directly proportional to the shear viscosity. In kinetic theory, however, the response is also sensitive to all other non-hydrodynamic modes in the $\ell=2$ channel owing to their nontrivial momentum-space structure.

\emph{Non-hydrodynamic sector, $\ell>2$}:
In the \(k\to0\) limit, the hydrodynamic description contains angular
structure only up to the viscous quadrupole sector,
\(\ell=0,1,2\). Higher angular moments therefore lie entirely outside
the hydrodynamic sector and directly probe non-hydrodynamic excitations
of the kinetic theory.
The jet wake therefore provides a direct spectroscopic probe of the
non-hydrodynamic sector of the quark-gluon plasma.

\begin{figure}
    \centering
\includegraphics[width=0.99\linewidth]{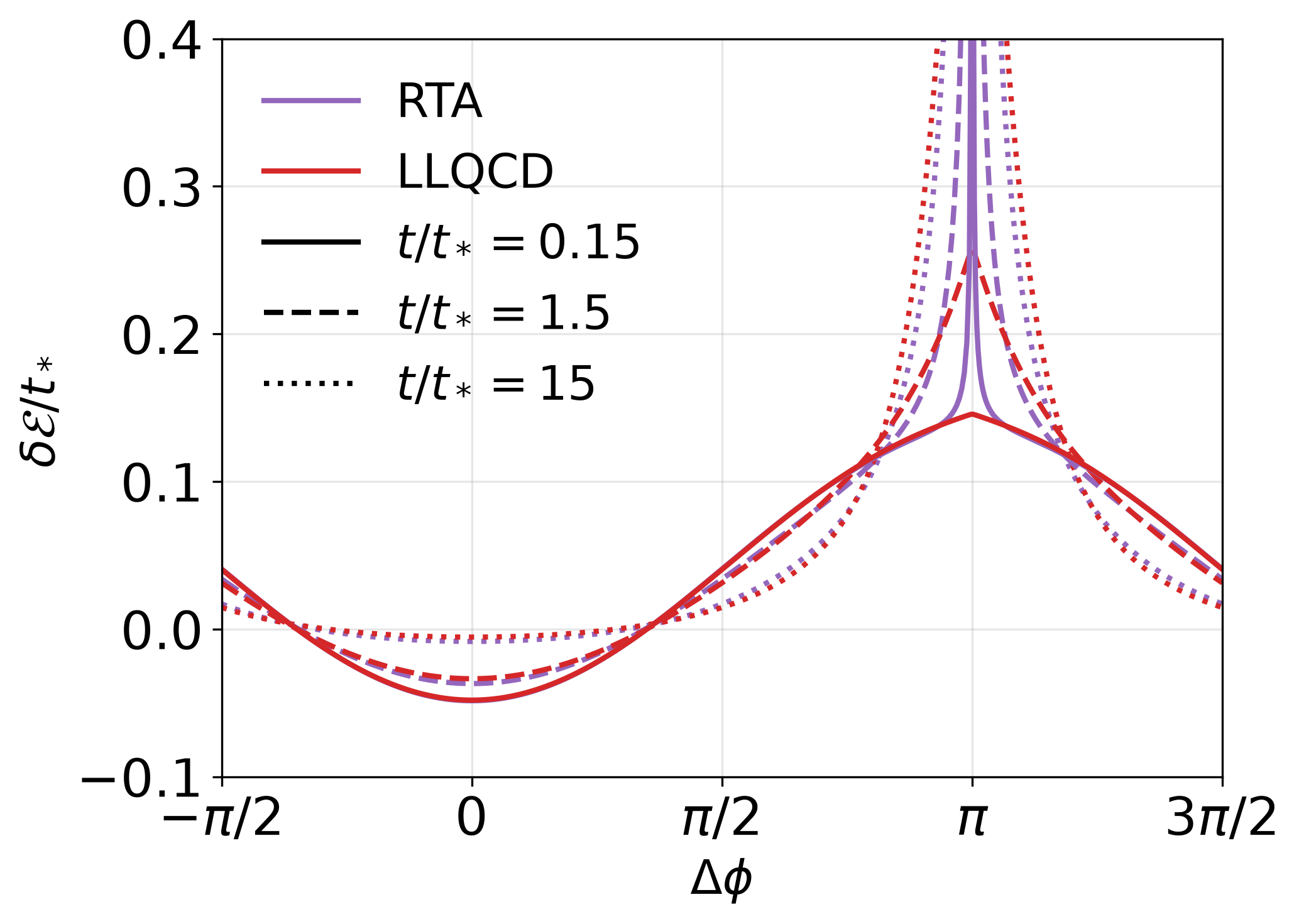}
    \caption{The evolution of the jet-wake image, defined in Eq.~(\ref{eq:image}), as a function of the freeze-out time $t_*$ for RTA and leading-log QCD kinetic theories. At later times, the hydrodynamic $\ell =0,1$ modes come to dominate as their contribution is proportional to $t_*$. For both models, we fix $\eta/s= 0.12$ and $v_{\rm jet}=1$.
    }
    \label{fig:image1}
\end{figure}

{\bf Sensitivity of detector moments.} We now illustrate that detector moments \eqref{eq:image} are sensitive to the microscopic dynamics underlying equilibration in the plasma. To this end, we compare three simple kinetic theories:  relaxation-time approximation (RTA), which propagates quasi-particles with a common relaxation time $\tau_R$; a diffusive dynamics where relaxation is governed by the angular Laplacian \(\Delta_\Omega\), and the perturbative (pure-glue) QCD in leading-log approximation \cite{Hong:2010at}, which contains a richer spectrum of non-hydrodynamic excitations and reproduces the QCD transport coefficients at leading logarithmic accuracy~\cite{Arnold:2000dr}.
The corresponding collision kernels are
\begin{align}
C_{\rm RTA}[f]
&=
\frac{p\cdot u}{\tau_R}
\left(
f-f_{\rm eq}
\right), \label{RTA}
\\
\int_{\bf p} E_p C_{\rm diff}[f]
&=
\frac{1}{6\tau_D}\,
\Delta_\Omega
\int_p E_p f,
\label{diff}
\\
C_{\rm LLQCD}[f]
&=
T\mu_A\,
\nabla_{\mathbf p}\cdot
\left[
\nabla_{\mathbf p} f
+
\frac{\hat{\mathbf v}_p}{T}\,f(1+f)
\right]. \label{LLQCD}
\end{align}
In all three cases, gain terms are added to enforce exact energy and momentum conservation~\cite{Hong:2010at}.

The spectrum of hydrodynamic and non-hydrodynamic modes in RTA kinetic theory has been discussed in~\cite{Romatschke:2015gic,Kurkela:2019kip,Heller:2018qvh,Bajec:2024jez,Bajec:2025dqm}. Diffusion kinetic theory corresponds to a Fokker--Planck-type evolution operator whose spectral decomposition has been widely studied in the mathematical and plasma-physics literature~\cite{Landau1936,Rosenbluth1957}.

A straightforward computation gives
\begin{equation} \tau_{n\ell}^{\rm RTA} = \tau_R, \qquad \tau_{n\ell}^{\rm diff} =
\frac{6\tau_D}{\ell(\ell+1)}, \qquad \ell\ge2,
\end{equation}
whereas the LLQCD relaxation times are obtained numerically. We choose
\(\tau_R=\tau_D\approx 5\times 0.461 /\mu_A \), such that all three models reproduce
the same shear viscosity,
$\eta/s
=
T\tau_R/5,
$
thereby fixing the hydrodynamic response while allowing the non-hydrodynamic relaxation spectrum to differ. 

The time evolution of the detector images of the jet wake obtained from the RTA and LLQCD kernels is shown in Fig.~\ref{fig:image1}. The two models produce distinct angular patterns, reflecting their different non-hydrodynamic relaxation spectra. The sensitivity to the nature of the non-hydrodynamic modes is particularly pronounced in the jet direction around $\Delta \phi = \pi$, and at sufficiently early times. In contrast, the negative jet wake opposite to the jet direction, around $\Delta \phi = 0$, emerges with nearly the same angular shape in both models, since it is largely dominated by the conserved hydrodynamic modes.

\begin{figure*}
    \centering
    \includegraphics[width=0.48\linewidth]{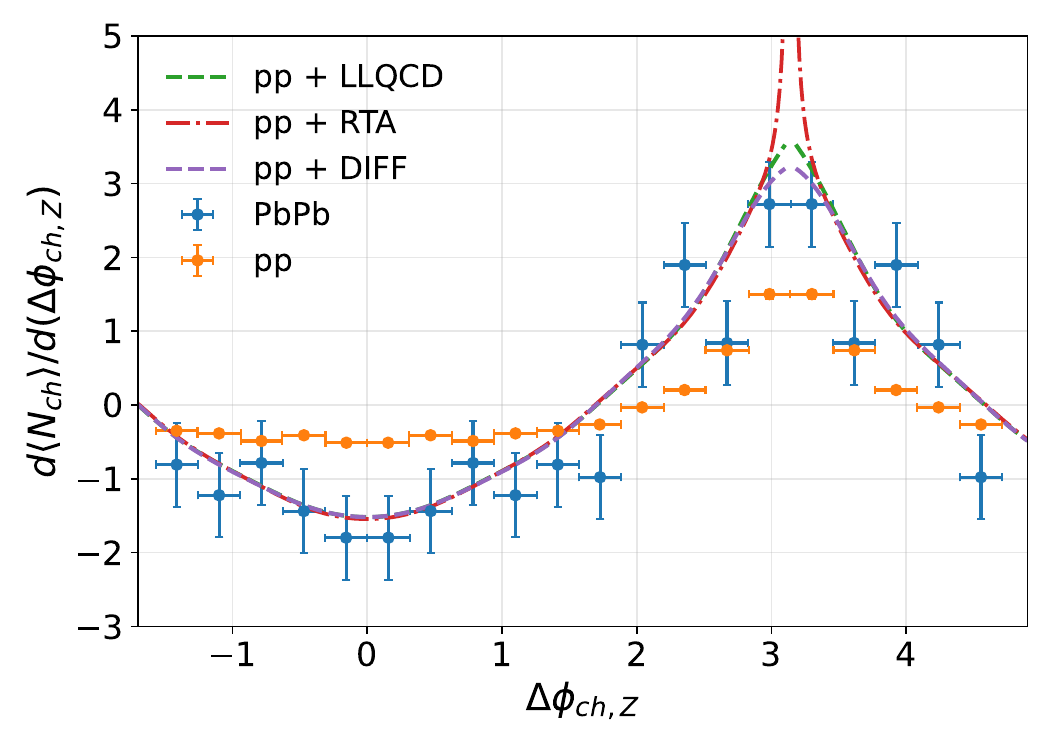}
    \includegraphics[width=0.48\linewidth]{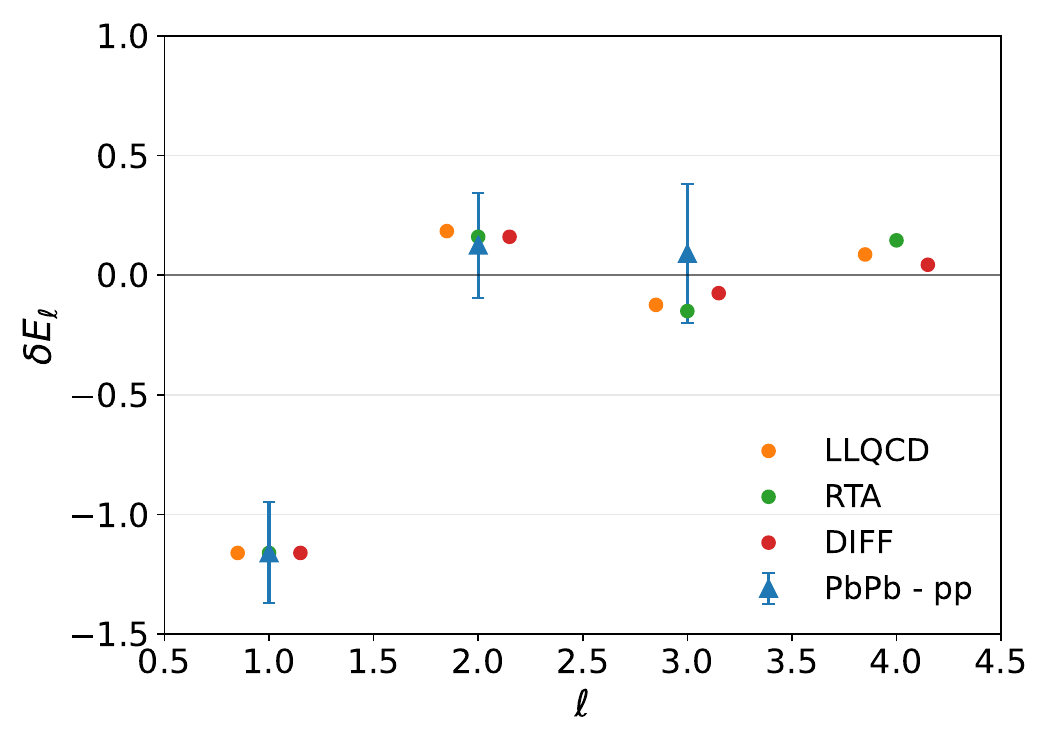}
    \caption{Negative jet-wake yield extracted from the kinetic theories defined in Eqs.~(\ref{RTA}--\ref{LLQCD}), compared with the CMS measurement of the charged particle yield as a function of the azimuthal angle $\Delta \phi$ from \cite{CMS:2025dua}. The negative jet wake is well described by all three models. For the models, we use $\eta/s = 0.12$ and $Tt_* = 4.5$ and the theoretical medium-induced modification is added to the measured pp distribution to obtain the PbPb prediction. The overall source normalization is fixed by requiring the resulting PbPb--pp difference to reproduce the measured $\ell=1$ mode.
    }
    \label{fig:comparison}
\end{figure*}

We have calculated the spherical harmonics for all three kinetic theory models. In each case, the jet wake exhibits distinct signatures of non-hydrodynamic behaviour, as evidenced by the non-vanishing $\ell>2$ harmonics (see Fig.~\ref{fig:comparison}). On general grounds, the lower moments with $\ell>2$ are the most promising candidates for a comparatively clean experimental probe of non-hydrodynamic modes, as they are independent of hydrodynamic modes while remaining least sensitive to the poorly controlled very forward jet fragmentation region that may affect higher moments.

The right-hand side of Fig.~\ref{fig:comparison} overlays the energy-flow moments calculated in the three kinetic models with the particle-flow moments extracted from the CMS data. Although these are, strictly speaking, different experimental observables, their comparison provides a qualitative indication of the current experimental precision, the improved accuracy required to establish a non-vanishing $\ell=3$ harmonic experimentally, and the still greater precision needed to distinguish between different microscopic models, which predict different higher harmonics.

{\bf Summary and outlook.}
The central message of this Letter is that the higher angular moments of the negative jet wake should become a target of precision experimental measurements, as they encode qualitatively new physics that has so far remained elusive in QGP phenomenology. While the lowest angular moments probe energy deposition and viscous transport, higher moments directly probe microscopic relaxation dynamics.
We have shown that precision measurements of these higher moments are uniquely sensitive to non-hydrodynamic modes: any non-zero measurement of a harmonic with $\ell \geq 3$ would therefore provide unambiguous evidence that the QGP is more than a hydrodynamic fluid.

Importantly, this conclusion is generic and does not rely on model-dependent assumptions. In particular, it is independent of the simplified kinetic models used here to illustrate this point. Establishing whether the higher moments of the jet wake deviate from the hydrodynamic null hypothesis -- and hence whether non-hydrodynamic modes are present in the QGP -- requires only sufficient experimental precision, not greater theoretical predictive power.

In contrast, improved theoretical modeling will be needed to determine the physical nature of any detected non-hydrodynamic modes. The three simple models studied here all predict small but distinctly non-zero harmonics with $\ell \geq 3$, whose amplitudes are consistently somewhat smaller than, but comparable to, that of the $\ell = 2$ mode. These results provide a first indication that the subtle yet conceptually important signatures of non-hydrodynamic modes may indeed lie within the reach of modern precision measurements of the negative jet wake.
The simplicity of these models also enabled us to establish a particularly transparent analytical connection between the jet wake and the spectrum of non-hydrodynamic modes. At the same time, however, this simplicity limits their quantitative predictive power, preventing, for example, quantitative statements about the expected differences between signatures arising from RTA quasi-particles and those associated with qualitatively different classes of non-hydrodynamic modes.

Turning this conceptual insight into a fully fledged phenomenological tool capable of distinguishing quantitatively between different non-hydrodynamic scenarios will require more realistic modeling. In the simplified models considered here, the jet source is represented as an eikonal color charge interacting with the medium through classical two-gluon exchange, whereas realistic jets undergo broadening, splitting, energy loss, and eventual thermalization. Likewise, the medium is treated as a static, homogeneous brick without longitudinal or transverse flow, while in realistic heavy-ion collisions both the expanding background and the evolving jet structure modify the detector image quantitatively. These and other effects can change the quantitative amplitudes of the higher harmonics and their relation to the underlying non-hydrodynamic modes. However, they do not affect the qualitative conclusion that harmonics with $\ell \geq 3$ vanish identically in hydrodynamics but become non-zero in the presence of non-hydrodynamic modes. Understanding these quantitative effects will require more detailed simulations, motivating a systematic exploration of realistic numerical simulations \cite{Yang:2022nei}, a more detailed incorporation of the angular structure of the jet wake in other phenomenological approaches \cite{Casalderrey-Solana:2014bpa}, or possibly the study of other classes of observables such as energy-energy correlators \cite{Moult:2025nhu,Hofman:2008ar}.

We conclude by reiterating that, on general grounds, the plasma of every relativistic quantum field theory necessarily supports non-hydrodynamic modes. The question is therefore not whether such modes are part of the quark--gluon plasma, but only whether the experimental and theoretical heavy-ion programme can develop sufficiently refined techniques to detect them unambiguously. We hope that our proposal to use the jet wake as a spectroscopic probe will contribute to this goal.

\section{Acknowledgments}
We thank Gabriel Cuomo, Pavel Kovtun, Juan Maldacena, Guy Moore, Berndt Mueller, and Krishna Rajagopal for useful discussions. We thank the KITP Santa Barbara for hospitality while this project was initiated (NSF PHY-2309135). 
A.K. is supported by the Research Council of Norway (CoreQCD, project number 361873). A.S. is supported by funding through the UL Startup project (UNLOCK) of the Slovenian Research Agency under contract no.~SN-ZRD/22-27/510. I.M. is supported by the Sloan Foundation.

\bibliographystyle{apsrev4-2}
\bibliography{references}

\appendix
\section{Appendix: Details on the jet source}

\begin{figure}
    \centering
    \includegraphics[width=1.0\linewidth]{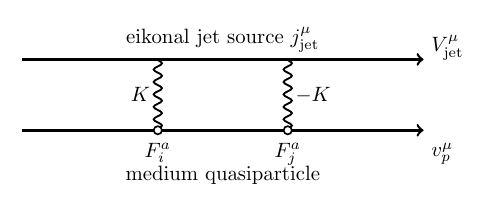}
    \caption{Diagram of the two-gluon exchange process giving rise to the source discussed in the Appendix.}
    \label{fig:placeholder}
\end{figure}

In this appendix we review the construction of the jet-induced source
term introduced in Ref.~\cite{Neufeld:2008fi}, adapted to the framework
used in this work. A hard color charge propagating through the medium at
fixed velocity \(v_{\rm jet}\)  generates
chromoelectric and chromomagnetic fields, which perturb the distribution
of medium quasiparticles through the Lorentz force. Because the equilibrium medium is color neutral, a single insertion of the jet-induced color field averages to zero. The leading color-singlet response therefore arises from terms quadratic in the fields, corresponding schematically to a two-gluon exchange with the jet (see Fig.~\ref{fig:placeholder}). The resulting color-singlet response can be written as a spacetime-dependent source term of Fokker--Planck form,
\begin{equation}
S(x,\mathbf p,t)
=
\nabla_{p_i}
\left[
D_{ij}(x,\mathbf p,t)\,
\nabla_{p_j} f_0(p)
\right],
\end{equation}
where \(D_{ij}\) is determined by a force-force correlator evaluated
along the trajectory of a medium particle,
\begin{equation}
D_{ij}(x,\mathbf p,t)
=
\int_0^\infty d\tau\,
F_i(x,t)\,
F_j(x-\mathbf v_p \tau, t-\tau).
\end{equation}

In Ref.~\cite{Neufeld:2008fi}, this construction was used to determine
the induced energy-momentum deposition current \(J^\mu(x)\) that sources
hydrodynamics. Here, instead, we use the same microscopic setup to
determine the angular moments of the kinetic source in the long-wavelength limit, relevant for the detector image.

In this limit, the source depends on the
particle momentum only through its magnitude \(p\) and its direction
relative to the jet axis,
\begin{equation}
\chi
\equiv
\hat{\mathbf p}\cdot\hat{\mathbf v}_{\rm jet}.
\end{equation}
Our goal is therefore to determine the spherical-harmonic moments
\begin{equation}
 S_\ell(p; v_{\rm jet})
=
\frac{2\ell+1}{2}
\int_{-1}^{1}
d\chi\,
P_\ell(\chi)\,
S(p, \chi; v_{\rm jet}),
\end{equation}
which control the excitation of the different angular channels of the
jet wake discussed in the main text.

Throughout this appendix we assume a homogeneous equilibrium background,
an infinitely long straight jet trajectory, and linear response. In
addition to the assumptions made in the main text, we further specialize
to free gauge propagators and massless medium quasiparticles.

The color current associated with an eikonally propagating point-like jet particle is given in Fourier space by
\footnote{Our convention for the Fourier transform is
$
f(x,t)
=
\int \frac{d^4K}{(2\pi)^4}
f(K)
e^{-ik^0 t + i\mathbf k\cdot \mathbf x}.
$
}
\begin{equation}\label{eq:jet-source}
j^{\mu,a}_{\rm jet}(K)
=
\frac{g\,Q_{\rm jet}^a}{\gamma}\,
V^\mu_{\rm jet}\,
(2\pi)\delta(k^0-\mathbf k\cdot \mathbf v_{\rm jet}),
\end{equation}
where \(Q_{\rm jet}^a\) denotes the color charge of the jet particle.
We align the jet along the \(z\)-direction,
\(V^\mu_{\rm jet}=\gamma(1,0,0,v_{\rm jet})\). 
The corresponding color-field perturbation is obtained by convolving the
jet current with the retarded gluon propagator, $D^{R}_{{\mu\nu}}$,  
\begin{equation}
\delta A_\mu^a(K)
=
D^R_{\mu\nu}(K)\,
j_{\rm jet}^{\nu,a}(K).
\end{equation}
The color-Lorentz force exerted by the color field \(\delta A_\mu^a\) on
a particle moving with velocity \(\mathbf v_p\) and color charge
\(Q_{\rm med}^a\) is
\begin{equation}
F_i(K)
=
g Q_{\rm med}^a F_i^a(K),
\,\, 
F_i^a(K)
\equiv
\mathcal V_i^{\mu}(K)\,
\delta A_\mu^a(K).
\end{equation}
Here \(\mathcal V_i^{\mu}\) captures the momentum dependence of the
Lorentz vertex, defined through
$
\mathcal V_i^{\mu}(K) A_\mu(K)
=
E_i(K)+(\mathbf v_p\times\mathbf B(K))_i.
$

Since the force is proportional to the medium-particle color charge,
the force-force correlator involves the product
\(Q_{\rm med}^aQ_{\rm med}^b\). Averaging over the color ensemble of a color-neutral
medium gives
\begin{equation}
\left\langle
Q_{\rm med}^a Q_{\rm med}^b
\right\rangle
=
\frac{C_2}{N_c^2-1}\delta^{ab},
\end{equation}
where \(C_2\) is the quadratic Casimir of the medium-particle
representation. This color average makes explicit that terms linear in
the jet-induced field average to zero in a color-neutral medium,
\(\langle Q_{\rm med}^a\rangle=0\), so that the leading color-singlet
response is quadratic in the field.

Passing to Fourier space, the color-averaged force-force correlator
takes the form
\begin{equation}\label{eq:dij}
D_{ij}(\mathbf p)
=
\frac{ig^2 C_2}{N_c^2-1}
\int \frac{d^4K}{(2\pi)^4}
\frac{F_i^a(K)\,
F_j^a(-K)}
{-(k^0-\mathbf k\cdot \mathbf v_p)+i0^+}.
\end{equation}

To evaluate Eq.~(\ref{eq:dij}), we decompose the retarded gluon
propagator in Coulomb gauge into longitudinal and transverse
components,
\begin{align}
D^R_{00}(K)
&=
\Delta_L,
\,\,
D^R_{ij}(K)
=
\Delta_T
\left(
\delta_{ij}
-\frac{k_i k_j}{\mathbf k^2}
\right),
\end{align}
which for a free gluon are
\begin{equation}\label{eq:free-prop}
\Delta_L(K)
=
\frac{1}{\mathbf k^2},
\, \,
\Delta_T(K)
=
\frac{1}{(k^0+i0^+)^2-\mathbf k^2}.
\end{equation}

Using
$$
\frac{1}{x+i0^+}
=
{\cal P}\frac1x
-i\pi\delta(x),
$$
the retarded denominator in Eq.~(\ref{eq:dij}) can be decomposed into
principal-value and delta-function contributions. Since the numerator is
even under \(\mathbf k\rightarrow-\mathbf k\), the principal-value term
vanishes upon integration. 

The product of the two eikonal currents generates a squared delta
function. This divergence is an artifact of the assumption that the jet
has propagated through the medium for an infinite time. Regulating the
source by a finite propagation time \(t_{\rm jet}\) gives
\[
(2\pi)^2\delta^2
\rightarrow
(2\pi)t_{\rm jet}\delta .
\]
After dividing by \(t_{\rm jet}\), Eq.~(\ref{eq:dij}) becomes
\begin{equation}
\frac{D_{ij}(\mathbf p)}{t_{\rm jet}}
=
\frac{g^2 C_2}{N_c^2-1}
\int \frac{d^3k}{(2\pi)^3}\,
\pi\,\delta(\Omega)\,
\Gamma_{ij},
\end{equation}
where
\begin{equation}
\Omega
\equiv
\mathbf k\cdot(\mathbf v_{\rm jet}-\mathbf v_p),
\qquad
K_*^\mu
\equiv
(\mathbf k\cdot\mathbf v_{\rm jet},\,\mathbf k).
\end{equation}
On the support of the delta functions, the color-resolved force product
reduces to
\begin{align}
\Gamma_{ij}
&=
g^2 C_{\rm jet}\,
k_i k_j
\Big[
\Delta_L(K_*)\Delta_L(-K_*) \nonumber
\\ -& 2\Lambda\,\Delta_L(K_*)\Delta_T(-K_*)
\nonumber
+\Lambda^2\,\Delta_T(K_*)\Delta_T(-K_*)
\Big],
\end{align}
where \(C_{\rm jet}\equiv Q_{\rm jet}^aQ_{\rm jet}^a\), and
\[
\Lambda
\equiv
\mathbf v_p\cdot\mathbf v_{\rm jet}^{\perp},
\qquad
\mathbf v_{\rm jet}^{\perp}
=
\mathbf v_{\rm jet}
-
\frac{\mathbf k\cdot\mathbf v_{\rm jet}}{\mathbf k^2}\,
\mathbf k,
\]
denotes the component of \(\mathbf v_{\rm jet}\) transverse to
\(\mathbf k\).

To obtain the scalar source \(S(p,\chi)\), we contract \(D_{ij}\) with
\(\nabla_{p_j}f_0=f_0'(p)\hat p_j\). It is useful to introduce the
orthonormal basis
\begin{equation}\label{eq:ortho-basis}
\hat{\mathbf e}_1=\hat{\mathbf p},
\qquad
\hat{\mathbf e}_2
\equiv
\hat{\mathbf e}_\perp
=
\frac{\hat{\mathbf v}_{\rm jet}-\chi\hat{\mathbf p}}
{\sqrt{1-\chi^2}},
\qquad
\hat{\mathbf e}_3
=
\frac{\hat{\mathbf p}\times\hat{\mathbf v}_{\rm jet}}
{\sqrt{1-\chi^2}} .
\end{equation}
Here \(\hat{\mathbf e}_1\) points along the particle momentum,
\(\hat{\mathbf e}_2\) lies in the plane spanned by
\(\hat{\mathbf p}\) and \(\hat{\mathbf v}_{\rm jet}\), and
\(\hat{\mathbf e}_3\) is orthogonal to that plane.

In this basis,
\begin{equation}
D_{ij}\hat p_j
=
D_{11}(p,\chi)\,\hat p_i
+
D_{12}(p,\chi)\,\hat e_{\perp i}.
\end{equation}

For free propagators, after performing the angular integrations, the
relevant components of the diffusion tensor become
\begin{align}
D_{11}^{\rm free}
&=
\mathcal{C} 
\frac{v_{\rm jet}^2(1-\chi^2)}
{\kappa^3}
I(m,h)
,
\\
D_{12}^{\rm free}
&=
\mathcal{C}
\frac{  v_{\rm jet}\sqrt{1-\chi^2}(v_{\rm jet}\chi-1)}
{\kappa^3}
I(m,h).
\end{align}
where
\begin{equation}
\kappa^2
=
1+v_{\rm jet}^2-2v_{\rm jet}\chi,
\qquad
m
=
\frac{v_{\rm jet}^2(1-\chi^2)}{\kappa^2},
\qquad
h
=
v_{\rm jet}\chi,
\end{equation}
and function containing the non-trivial part of the angular structure
\begin{align}
I(m,h)
&=
\frac{\pi}
{m(1-m)^{3/2}}
\Big[
h^2m
-10hm
+8h
+9m
-8
\nonumber\\
&
+
\left(
8hm-8h-4m^2-4m+8
\right)
\sqrt{1-m}
\Big].
\end{align}
The overall normalization is given by
\begin{equation}
\mathcal{C}
\equiv
 \frac{g^4 C_2 C_{\rm jet}}{N_c^2-1}
\left[
\int_0^\infty \frac{dk}{k}
\right].
\end{equation}
The logarithmic divergence originates from the long-range Coulomb
interaction of the free propagator. In a realistic medium, the infrared
region is regulated by screening effects, while the ultraviolet region
is cut off when the exchanged momentum becomes comparable to the typical
momentum of the medium quasiparticles and the soft-scattering
approximation ceases to apply. Here we leave these
effects implicit and retain only the resulting Coulomb logarithm.

The diffusion tensor depends on the particle momentum only through the
relative angle \(\chi\). Consequently, the source factorizes into two
angular structures with distinct radial weights,
\begin{equation}
S(p,\chi;v_{\rm jet})
=
R^{\parallel}(p)\,
\mathcal S^{\parallel}(\chi;v_{\rm jet})
+
R^{\perp}(p)\,
\mathcal S^{\perp}(\chi;v_{\rm jet}),
\end{equation}
where
\begin{align}
R^{\parallel}(p)
&=
\mathcal C \,
\frac{1}{p^2}
\frac{d}{dp}
\left[
p^2 f_0'(p)
\right], \quad
R^{\perp}(p)
=
\mathcal C \,
\frac{f_0'(p)}{p},
\end{align}
and
\begin{align}
\mathcal S^{\parallel}(\chi;v_{\rm jet})
&=
\frac{v_{\rm jet}^2(1-\chi^2)}
{\kappa^3}
I(m,h),
\\
\mathcal S^{\perp}(\chi;v_{\rm jet})
&=
-\frac{\partial}{\partial\chi}
\left[
\frac{
v_{\rm jet}(1-v_{\rm jet}\chi)
(1-\chi^2)
}{
\kappa^3
}
I(m,h)
\right].
\end{align}

The angular structures can be expanded in Legendre polynomials
\begin{equation}
\mathcal S^{\parallel,\perp}(\chi;v_{\rm jet})
=
\sum_{\ell=0}^{\infty}
\mathcal S^{\parallel,\perp}_{\ell}(v_{\rm jet})
P_\ell(\chi),
\end{equation}
The first few Legendre moments can be evaluated analytically. Defining the rapidity of the jet velocity
\begin{equation}
L\equiv \operatorname{arctanh}v_{\rm jet},
\end{equation}
one finds
\begin{align}
\mathcal S_{\parallel,0}
&=
\pi\left[
1+\frac{v_{\rm jet}^2-1}{v_{\rm jet}}L
\right],
\\
\mathcal S_{\parallel,1}
&=
\pi\left[
\frac{3}{v_{\rm jet}}
-2v_{\rm jet}
+\frac{8}{5}v_{\rm jet}^3
+
3\left(1-\frac{1}{v_{\rm jet}^2}\right)L
\right],
\\
\mathcal S_{\parallel,2}
&=
\frac{5\pi}{2}\left[
-3+\frac{3}{v_{\rm jet}^2}
+\frac{48}{35}v_{\rm jet}^4
-
\frac{3-4v_{\rm jet}^2+v_{\rm jet}^4}{v_{\rm jet}^3}L
\right],
\\
\mathcal S_{\parallel,3}
&=
\frac{\pi}{30v_{\rm jet}^4}
\Big[
v_{\rm jet}
\big(
525-665v_{\rm jet}^2
+140v_{\rm jet}^4
-48v_{\rm jet}^6
\nonumber\\
&
+160v_{\rm jet}^8
\big)
-
105
\left(
5-8v_{\rm jet}^2+3v_{\rm jet}^4
\right)L
\Big].
\end{align}

\begin{align}
\mathcal S_{\perp,0}
&=
0,
\\
\mathcal S_{\perp,1}
&=
%-
\frac{2\pi}{5}v_{\rm jet}
\left(5-4v_{\rm jet}^2\right),
\\
\mathcal S_{\perp,2}
&=
%-
\frac{8\pi}{7}v_{\rm jet}^2
\left(7-6v_{\rm jet}^2\right),
\\
\mathcal S_{\perp,3}
&=
%-
\frac{8\pi}{5}v_{\rm jet}^3
\left(11-10v_{\rm jet}^2\right).
\end{align}
The moments \(\mathcal S^{\perp}_{\ell}\) are simple polynomials in
\(v_{\rm jet}\), whereas the nontrivial logarithmic dependence on the
jet velocity resides entirely in \(\mathcal S^{\parallel}_{\ell}\).

The lowest angular moments are fixed by the total energy and momentum
deposited into the medium. Since the source \(S(\mathbf p)\) is derived
for a single spin-color degree of freedom of a medium quasiparticle, the
physical energy-momentum deposition current is obtained by summing over
the corresponding degeneracy,
\begin{equation}
J^\mu({\bf k} =0 )
=
\nu_s
\int \frac{d^3p}{(2\pi)^3}\,
p^\mu S(\mathbf p).
\end{equation}
This projection receives contributions only from the
\(\ell=0\) and \(\ell=1\) moments of the source. Using the expressions
for \(\mathcal S_0\) and \(\mathcal S_1\) above, one finds the
collisional energy-momentum deposition
\begin{align}
\frac{dE}{dt}
&=
\frac{\nu_s g^4 C_2 C_{\rm jet}T^2}
     {24\pi (N_c^2-1)}
\left[
1+
\left(
v_{\rm jet}
-\frac{1}{v_{\rm jet}}
\right)
L
\right]
\ln\frac{T}{m_D},
\\
\frac{dP^z}{dt}
&=
\frac{1}{v_{\rm jet}}\,\frac{dE}{dt}.
\end{align}
The logarithm arises from integrating over momentum transfers between
the Debye screening scale \(m_D\) and the hard thermal scale \(T\). For
the pure-glue plasma,
\(\nu_s=\nu_g=2(N_c^2-1)\) and \(C_2=C_A=N_c\).

These results agree exactly, including the overall normalization, with
the integrated source current of Neufeld \emph{et al.}~\cite{Neufeld:2008fi}.
The same coefficient is obtained in the leading-logarithmic energy-loss
calculation of Moore and Teaney~\cite{Moore:2004tg}, where the drag
force of a heavy quark is computed from quantum QCD scattering
amplitudes.
%\newpage

In the
\(v_{\rm jet}\rightarrow1\) limit its angular dependence approaches a
universal limiting form. This limiting shape develops an integrable
forward enhancement,
\[
S(p,\chi;v_{\rm jet}\to1)
\sim
\frac{1}{\sqrt{1-\chi}}
\sim
\frac{1}{\theta},
\]
in the direction of the jet. This forward singularity is integrable and all
Legendre moments remain finite in the ultrarelativistic limit.

Finally, the same elementary force-force correlator determines the
leading-logarithmic transverse momentum broadening coefficient of an
energetic jet. Averaging over the thermal ensemble of medium
quasiparticles, one obtains
\begin{equation}
\hat q
=
2P_\perp^{ij}\,
\nu_s
\int\frac{d^3p}{(2\pi)^3}\,
f_0(p)\left[1+f_0(p)\right]\,
D_{ij}(\mathbf p),
\end{equation}
where
\[
P_\perp^{ij}
=
\delta^{ij}
-
\hat v_{\rm jet}^i\hat v_{\rm jet}^j
\]
projects onto the directions transverse to the jet.
Using the same free propagators as above, this reproduces the standard
leading-logarithmic result
\begin{equation}
\hat q^{\rm LL}
=
\frac{g^4 C_{\rm jet} C_2 T^3}{6\pi}
\ln\frac{T}{m_D},
\end{equation}
Since both $\hat q$ and the source moments are determined by the same
force-force correlator, they share the common normalization
$\mathcal C$. Consequently,
\begin{equation}
\mathcal C
=
\frac{6\pi}{N_c^2-1}
\frac{\hat q}{T^3},
\end{equation}
allowing the overall normalization of the source to be expressed
directly in terms of the phenomenological value of $\hat q/T^3$.

\end{document}